\documentclass[aps,prd,preprint,superscriptaddress,tightenlines,nofootinbib]{revtex4}
\usepackage{graphicx} 
\usepackage{dcolumn}
\usepackage{bm}
\usepackage{epsfig}

\newcommand{\Y}{\Upsilon}
\newcommand{\ra}{\rightarrow}

\begin{document}

\preprint{CLNS 05-1934}       
\preprint{CLEO 05-22}

\title{ Observation of $B_s$ Production 
                      at the $\Upsilon(5S)$ Resonance  }

\author{G.~Bonvicini}
\author{D.~Cinabro}
\author{M.~Dubrovin}
\author{A.~Lincoln}
\affiliation{Wayne State University, Detroit, Michigan 48202}
\author{A.~Bornheim}
\author{S.~P.~Pappas}
\author{A.~J.~Weinstein}
\affiliation{California Institute of Technology, Pasadena, California 91125}
\author{D.~M.~Asner}
\author{K.~W.~Edwards}
\affiliation{Carleton University, Ottawa, Ontario, Canada K1S 5B6}
\author{R.~A.~Briere}
\author{G.~P.~Chen}
\author{J.~Chen}
\author{T.~Ferguson}
\author{G.~Tatishvili}
\author{H.~Vogel}
\author{M.~E.~Watkins}
\affiliation{Carnegie Mellon University, Pittsburgh, Pennsylvania 15213}
% ...
\collaboration{CLEO Collaboration} 
\noaffiliation

\date{October 12, 2005}

\begin{abstract} 

Using the CLEO detector at the Cornell Electron Storage Ring,  we have 
observed the $B_s$ meson in $e^+ e^-$ annihilation at the $\Y(5S)$ 
resonance. We find 14 candidates consistent with $B_s$ decays into 
final states with a $J/\psi$ or a $D_s^{(*)-}$. 
The probability that we have observed a background fluctuation is 
less than $8 \times 10^{-10}$. We have established that at the 
energy of the  $\Y(5S)$ resonance $B_s$ production proceeds 
predominantly through the creation of $B_s^{*} \bar{B}_s^{*}$ pairs. 
We find $\sigma(e^+ e^- \ra B_s^* \bar{B}_s^*) 
= [0.11^{+0.04}_{-0.03}{\rm (stat.)} \pm 0.02 {\rm (syst.)}]  \; {\rm nb}$,
and set the following limits:
$\sigma(e^+ e^- \ra  B_s \bar{B}_s) /
\sigma( e^+ e^- \ra B_s^{*} \bar{B}_s^{*}) < 0.16 $ 
and 
$[\sigma(e^+ e^- \ra B_s \bar{B}_s^{*}) + 
\sigma(e^+ e^- \ra  B_s^{*} \bar{B}_s ) ]/
\sigma( e^+ e^- \ra B_s^{*} \bar{B}_s^{*} ) 
< 0.16 $~(90\%~CL). 
The mass of the $B_s^*$ meson is measured to be 
$M_{B_s^*} = [ 5.414 \pm 0.001{\rm (stat.)} 
                      \pm 0.003{\rm (syst.)}]$~GeV/$c^2$.

\end{abstract}

\maketitle

The $\Y(5S)$ resonance was discovered by the CLEO and CUSB 
collaborations~\cite{CLEOY5S}. It lies about 
40 MeV above the $B_s^* \bar{B}_s^*$ production threshold. 
At this energy $B_{(s)}$ mesons can be produced in a variety 
of states $B_{(s)}^{(*)} \bar{B}_{(s)}^{(*)} (\pi) (\pi)$. 
The cross section in this energy region is well described 
by the Unitarized Quark Model~\cite{theory},
which predicts that the total $b \bar{b}$  cross section, 
measured to be about 0.35~nb~\cite{CLEOY5S}, is dominated 
by $B_{(s)}^*\bar{B}_{(s)}^*$  production with $B_s^*\bar{B}_s^*$ 
constituting one third of it. Knowledge of the $B_s$ production 
mechanism and rate at the $\Y(5S)$ resonance is essential for 
evaluating the physics potential of the $B_s$ program at a 
future $e^+ e^-$ Super-$B$ Factory~\cite{superb}.

In this Letter, we report the first observation of fully
reconstructed $B_s$ mesons produced in $e^+ e^-$ annihilation 
at the energy of the $\Y(5S)$ resonance. 
We demonstrate for the first time that the dominant $B_s$ 
production mechanism at this energy is $B_s^* \bar{B}_s^*$,
measure the cross section for this process, set upper 
limits on the competing mechanisms, and thereby test 
theoretical predictions. 
A companion Letter using the same data set 
reported first evidence for $B_s^{(*)} \bar{B}_s^{(*)}$ 
production from a measurement of the 
$D_s^+$ inclusive yield~\cite{y5sinclusive}.

An extension of the exclusive $B$ meson reconstruction 
technique used at the $\Y(4S)$ resonance is employed to 
reconstruct $B_s$ mesons at the $\Y(5S)$~\cite{ourtechnique}. 
Signal events are identified in the search plane 
of two variables: 
$M_{\rm bc} \equiv \sqrt{E_{\rm beam}^2/c^4 - |\vec{p}_{B_s}|^2/c^2}$ 
and 
$\Delta E  \equiv E_{ B_s } - E_{\rm beam}$. The signal regions 
in the search plane are chosen using 
$M_{B_s} = (5.3660 \pm 0.0008)$~GeV/$c^2$~\cite{bsmass} and 
$M_{B_s^*} -  M_{B_s}= (47.0 \pm 2.6)$~MeV/$c^2$~\cite{pdg}. 
We assume that the $B_s^*$ meson decays to a $B_s$
meson via the emission of a 47~MeV photon with a branching
fraction equal to unity.

At the energy of the $\Y(5S)$ resonance, the following 
states containing a $b \bar{s}$ quark pair are possible:
$B_s \bar{B}_s$, $B_s \bar{B}_s^*$ (or $B_s^* \bar{B}_s$),
and $B_s^* \bar{B}_s^*$. If the production of $B_s$ mesons 
occurs through the creation of  $B_s \bar{B}_s$ pairs, 
$M_{\rm bc} = M_{B_s}$ and $\Delta E = 0$~MeV.
In the two cases involving $B_s^*$ production, to increase
the reconstruction efficiency, the soft photon from 
the $B_s^*$ meson is not reconstructed. This leads to a shift 
from zero in $\Delta E$ but negligible smearing of the 
$B_s$ momentum, as the photon carries a small fraction 
of the total $B_s^*$ momentum.
For $B_s^* \bar{B}_s^*$, $E_{B_s}$ tends to be $47$~MeV smaller 
than $E_{\rm beam}$~($\Delta E = M_{B_s} - M_{B_s^*}$), 
and $M_{\rm bc}$ is $47$~MeV/$c^2$ higher than $M_{B_s}$
($M_{\rm bc} = M_{B_s^*}$), because the $47$~MeV photon 
is not reconstructed. If $B_s$ mesons are produced via 
$B_s \bar{B}_s^*$ or $B_s^* \bar{B}_s$ pair creation,
$M_{\rm bc}$ and $\Delta E$, to a good approximation, 
peak at $\frac{1}{2}  (M_{B_s} + M_{B_s^*})$ 
and $\frac{1}{2} (M_{B_s} - M_{B_s^*})$, respectively.
We define a signal band in the search plane as  
$- 60{\rm \;MeV} \le  \Delta E + [M_{\rm bc} - M_{B_s}]c^2 
\le +60{\rm \;MeV}$. 
Within the signal band, there are three signal regions,
each  about 24~MeV/$c^2$ wide and centered at  
$M_{\rm bc} =  $ $5.366$, $5.390$ and 
$5.413$~GeV/$c^2$ corresponding to $B_s \bar{B}_s$,  
$B_s \bar{B}_s^*$ and $B_s^* \bar{B}_s^*$ production, 
respectively. Identical signal regions are used for all 
$B_s$ modes, each corresponding to about 3~standard 
deviations (3$\sigma$) in $M_{\rm bc}$ and 
2 to 4$\sigma$ in $\Delta E$, depending on the mode.

The data used in this analysis were recorded by 
the CLEO~III detector at Cornell Electron Storage Ring~(CESR). 
CLEO~III  is  a general multipurpose solenoidal detector
designed to provide excellent charged and neutral particle
reconstruction efficiency and resolution. It has been described 
in detail in Ref.~\cite{cleodetector}.
The integrated luminosity of the data sample collected 
in the vicinity of the $\Y(5S)$ peak is 0.42~fb$^{-1}$,
most of which was taken at a center-of-mass energy  
$E_{\rm CM} = (10.859 \pm 0.006)$~GeV. 
A data sample of 7.6~fb$^{-1}$ 
collected at, and just below, the $\Y(4S)$ resonance and a data sample 
of 0.7~fb$^{-1}$ collected at $11.2~{\rm GeV} < E_{\rm CM} < 11.4~{\rm GeV}$ 
($\Lambda_b^0$-scan data)~\cite{LambdaB} is used to study background from $B$ 
mesons and continuum events of the type $e^+ e^- \rightarrow q \bar{q}$, 
where $q$ is $u$, $d$, $s$ or $c$~quark.

Tracks and showers used in reconstruction must satisfy a set of 
quality criteria. Primary tracks must be in the fiducial volume of 
the detector, come from the interaction point and have momenta above 
50~MeV/$c$. Identification of hadrons utilizes measurements of $dE/dx$ 
and information from a Ring Imaging Cherenkov Detector~(RICH). 
Pion or kaon candidates are required to have $dE/dx$ measurements 
within 3.0$\sigma$ of the expected value, 
and for tracks with momenta greater than 700~MeV/$c$, RICH information, 
if available, is combined with $dE/dx$ information.  
Electrons are identified above 700~MeV/$c$ using the ratio of the 
energy deposited in the calorimeter to the track momentum,
and $dE/dx$ information. Muon identification is efficient above 
1.0~GeV/$c$ and is based on the information from the muon chambers 
and the energy associated with the track in the calorimeter.

Each shower must not be matched to a track or be consistent 
with a hadronic fragment. 
The shower cannot be associated with noisy crystals in the
calorimeter and its energy must be greater than 30~MeV. 
Neutral pion candidates are selected from pairs of photons 
with invariant mass within $2.5 \sigma$ 
($\sigma \sim 6.0$~MeV/$c^2$) of the $\pi^0$ mass. 
A mass constraint is used for $\pi^0$ candidates 
to improve their energy resolution in further 
reconstruction.

$B_s$ mesons are reconstructed in modes with a $J/\psi$ or 
a $D_s^{(*)-}$ meson. 
(Charge-conjugate modes are implied throughout this Letter.) 
We describe each of these in turn.
The following modes with a $J/\psi$ are reconstructed:
$J/\psi \phi$, $J/\psi\eta  $ and $J/\psi \eta^\prime$, 
where $\phi$, $\eta$ and $\eta^\prime$ mesons are reconstructed 
using $\phi \rightarrow K^+ K^-$, $\eta \rightarrow \gamma \gamma$ 
and $\eta^\prime \rightarrow \eta(\gamma \gamma) \pi^+ \pi^-$.

Two oppositely charged electron or muon candidates are combined to
form a $J/\psi$ candidate. In the reconstruction of $J/\psi 
\rightarrow e^+ e^-$, bremsstrahlung photons are recovered by 
using showers that are not matched to a track, but that line 
up with one or the other electron momentum vector within a 0.10~radian angle. 
The $J/\psi  \rightarrow \mu^+ \mu^-$ candidates are required to be 
within 35~MeV/$c^2$~(3.0$\sigma$) of the $J/\psi$ mass. The invariant 
mass window for  $J/\psi \rightarrow e^+ e^-$ is wider and asymmetric 
due to bremsstrahlung. Combinations satisfying  
$[M(e^+e^-) - M_{J/\psi}] \in [-150; 50]$~MeV/$c^2$ 
are accepted for further analysis.

We form $\phi$ candidates from pairs of oppositely charged 
tracks that do not satisfy stringent particle identification 
criteria for pions. The $\phi$ candidates within 
10~MeV/$c^2$ of the known $\phi$ mass are accepted. 
The $\eta$ candidates are formed from pairs of photons, 
each having an energy of at least 50 MeV, with an invariant mass 
within $2.5 \sigma$  of the known $\eta$ mass.  
A mass constraint is used for $\eta$ candidates in further reconstruction.
To reduce background from low energy photons and noise in the calorimeter, 
we require $\cos \theta_\gamma > -0.95$,
where $\theta_\gamma$ is the angle between the $\eta$ momentum vector 
in the laboratory frame and the momentum vector of the lower energy 
photon in the $\eta$ rest frame. The reconstruction of $\eta^\prime$ 
candidates is achieved by combining an $\eta$ candidate with any two 
oppositely charged tracks interpreted as pions and requiring the 
invariant mass of the combination be within 12~MeV/$c^2$ of the known 
$\eta^\prime$ mass. The $J/\psi$ is combined with a $\phi$, $\eta$ or 
$\eta^\prime$ candidate to form a $B_s$ candidate. If there are multiple 
$B_s$ candidates in an event, the candidate having the smallest distance 
to the center of the signal band along the $\Delta E$ axis is selected 
for each $B_s$ mode.

These selection criteria allow $B_s$ reconstruction with a very 
large signal-to-background ratio. We use data collected in the vicinity 
of the $\Y(4S)$ resonance and the $\Lambda_b^0$-scan data to study 
background. To correct for the difference in the beam energy between
these data and the $\Y(5S)$ data, 
$M_{\rm bc}$ is obtained using $\frac{E_{\Y(5S)}}{E_{\rm beam}} 
\sqrt{E_{\rm beam}^2/c^4 - |\vec{p}_{B_s}|^2/c^2}$.
The background shows no tendency to peak in the signal band. 
It decreases  with increasing $\Delta E$, and is approximately 
uniformly distributed throughout most of the $M_{\rm bc}$ 
range, tending to zero at the phase space limit $M_{\rm bc} = E_{\rm CM} / 2 $. 
The total number of non-$B_s$ background events in the entire search 
plane in the $\Y(5S)$ data is estimated to be $2.4 \pm 0.4({\rm stat.})$.
Backgrounds are well determined as the integrated luminosity of 
the background samples is a factor of twenty greater than 
that of the $\Y(5S)$ data sample.

\begin{figure}
\begin{center}
\epsfig{file=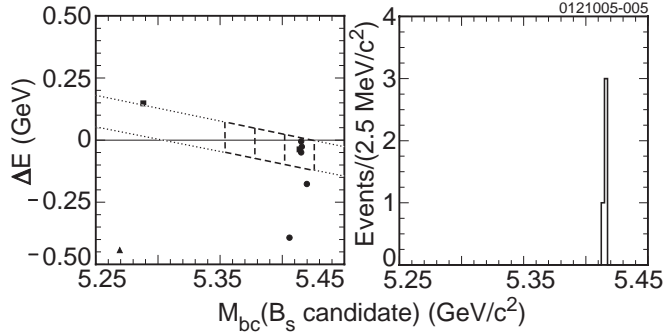,width=3.5in}
\end{center}
\caption{ The search plane~(left) and its projection on  $M_{\rm bc}$~(right)
          for events in the signal band  for $B_s$
          modes with a $J/\psi$ in the $\Y(5S)$ data.
          The circles, triangles and squares represent 
          $B_s \ra J/\psi \phi$, $J/\psi \eta$ and 
          $J/\psi \eta^\prime$ candidates, respectively.
          In the signal band (dotted lines),
          assuming $M_{B_s^*} -  M_{B_s}= 47$~MeV/$c^2$~\cite{pdg},
          $B_s$ mesons are expected to cluster within 
          the signal boxes (dashed lines) at
          $(5.366, 0.000)$, 
          $(5.390, -0.024)$ and 
          $(5.413, -0.047)$ for 
          $B_s \bar{B}_s$,  $B_s \bar{B}_s^*$ and $B_s^* \bar{B}_s^*$
          production, respectively.
        }
\label{jpsi_y5s_signal}
\end{figure}

Figure~\ref{jpsi_y5s_signal}
shows the search plane in the data~(left) and its projection on 
$M_{\rm bc}$~(right) for events in the signal band. There are 4 
events in the signal region all corresponding to $B_s^* \bar{B}_s^*$ 
production: 2~events in the $B_s \ra J/\psi(\mu^+ \mu^-) \phi$ mode, 
1~event in the $B_s \ra J/\psi(e^+ e^-) \phi$ mode, and 1~event in 
the $B_s \ra J/\psi(\mu^+ \mu^-) \eta^\prime$ mode. The rest of the 
search plane contains 4~background events: 2~events in the 
$B_s \ra J/\psi \phi$ mode, 1~event in the $B_s \ra J/\psi \eta$ 
mode, and 1~event in the  $B_s \ra J/\psi \eta^\prime$ mode.

To calculate the probability, $P_{\rm I}$, for the 
background to account for all events in the  $B_s^* \bar{B}_s^*$ 
signal region requires assumptions 
about the background shape in the search plane. 
To obtain a conservative estimate, we assume that the 
background density is uniform over the lower half 
of the search plane. 
The number of non-$B_s$ background events in the 
$B_s^* \bar{B}_s^*$ signal region is estimated from
the background study
to be less than 0.08 events at 68\%~Confidence Level~(CL). 
The Poisson probability for 0.08 background events to 
fluctuate to 4 or more events in the signal region is $P_{\rm I} = 
1.6 \times 10^{-6}$.

We now describe the analysis of $B_s$ modes with a $D_s^{(*)-}$ 
meson in the final state. 
As the $B_s$ is expected to decay almost 100
percent of the time to a $D_s^-$, these modes
provide access to a large fraction of $B_s$
decays. However, background from continuum 
production is significant and consequently stringent 
background suppression criteria must be applied.
We reconstruct the modes 
$\bar{B}_s \ra D_s^{(*)+} \pi^-$ and $\bar{B}_s \ra D_s^{(*)+} \rho^-$, 
where  the $D_s^+$ meson is reconstructed in the final states 
$K^+ K^0_S(\pi^+ \pi^-)$, 
$K^+ K^{*0}(K^- \pi^+)$, $\phi(K^- K^+) \pi^+$ and 
$\phi(K^- K^+) \rho^+(\pi^+ \pi^0)$, and the $D_s^{*+}$ meson 
is reconstructed in the $D_s^+ \gamma$ channel.

Hadron identification information is used only for kaons. 
The $\phi$/$K^{*0}$/$\rho^+$ candidates are constructed 
from ($K^-$ and $K^+$)/($K^-$ and $\pi^+$)/($\pi^0$ and 
$\pi^+$) candidates within 8/75/100~MeV/$c^2$ of their 
known mean masses, respectively. The $K^0_S$ candidates are 
formed from pairs of oppositely charged and vertex-constrained 
tracks, if the invariant mass is within 8 MeV/$c^2$ of the known 
$K^0_S$ mass, and the vertex is displaced from the beam interaction 
point by at least 3.0~mm. All $D_s^+$ candidates with an invariant 
mass within 3.0$\sigma$ of the known $D_s^+$ mass are used in 
further reconstruction.

The $D_s^{*+}$ candidates are reconstructed by combining 
the $D_s^+$ candidates with photons. The photon candidates 
are required to have energies, $E_\gamma$, in the kinematically allowed range:
$60~{\rm MeV} < E_\gamma < 400$~MeV. The mass difference $(M_{D_s^{*+}} - M_{D_s^+})$ 
is required to be within 2.0$\sigma$~($\sigma \sim 6$~MeV/$c^2$) 
of the known value in order to suppress 
a large background from random photons.

The $D_s^{(*)+}$ candidates are combined with a $\pi^-$
or a $\rho^-$. The $\rho^-$ candidates are required to have 
invariant mass within 100~MeV of the known mean value. 
We also require that the momentum of $\pi^0$ mesons 
from the $\rho^-$ candidates be above 200~MeV/$c$ to remove 
a large background in $\pi^0$ reconstruction at lower momenta.

In reconstruction of the decay sequences 
$P_i \ra V_f P_f$ with $V_f \ra p_1 p_2$, where 
$P$ or $p$ is a pseudoscalar and $V$ is 
a vector, the distribution of $\cos{\theta_{V}}$,
where $\theta_{V}$ is the angle between 
the $p_1$ momentum in the $V_f$ rest frame and the 
$V_f$ momentum in the $P_i$ rest frame, 
is proportional to $\cos^2{\theta_{V}}$, while 
the background tends to be uniform in this variable.
Accordingly, we require $|\cos{\theta_{V}}| > 0.60$ in the reconstruction  
of $D_s^+ \ra \bar{K}^{*0}(K^- \pi^+) K^+$, 
$D_s^+ \ra \phi(K^- K^+) \pi^+$ and $\bar{B}_s \ra D_s^+ \rho^-(\pi^- \pi^0)$.
Similarly, in the reconstruction of $\bar{B}_s \ra D_s^{*+} \pi^-$, 
we require $|\cos \theta_{\rm \gamma}| < 0.70 $, where
$\theta_{\rm \gamma}$ is the angle between the photon momentum 
in the $D_s^{*+}$ frame and the $D_s^{*+}$ momentum in the $\bar{B}_s$ 
frame. The distribution of $\cos \theta_{\rm \gamma}$  is proportional to 
$(1 - \cos^2 \theta_{\rm \gamma})$ for signal decays, while the 
background gradually increases towards $\cos \theta_{\rm \gamma} = -1$.

To suppress the continuum background the ratio of 
Fox-Wolfram moments $H_2$ and $H_0$~\cite{r2} is required 
to be less than 0.30. The continuum background is suppressed 
further using a requirement of $|\cos{\theta_{\rm thrust}}| < 0.70$, 
where $\theta_{\rm thrust}$ is the angle between the thrust axis of 
the $B_s$ candidate and the thrust axis of the rest of the event.

If there are multiple candidates in an event satisfying 
all selection criteria, we select one candidate 
for $\bar{B}_s \ra  D_s^+ \pi^-/\rho^- $ modes
and one candidate for $\bar{B}_s \ra  D_s^{*+} \pi^-/\rho^- $ 
modes. In each case the candidate with the smallest 
$ |\Delta E|^{\rm Signal\;Band} /  \sigma(\Delta E) $
is selected, where $|\Delta E|^{\rm Signal\;Band}$ is the distance to 
the center of the signal band along the $\Delta E$ axis and $\sigma(\Delta E)$ 
is mode dependent.

The same data samples as those in the analysis of $B_s$
modes with a $J/\psi$ are used in a background study. 
Again, the background shows no tendency to peak in the signal band,
and is similar in shape to the background in 
$B_s \rightarrow J/\psi \phi/ \eta / \eta^\prime$.
The total number of non-$B_s$ background events in the 
entire search plane in the $\Y(5S)$ data is estimated to 
be $47 \pm 2({\rm stat.})$.

\begin{figure}
\begin{center}
\epsfig{file=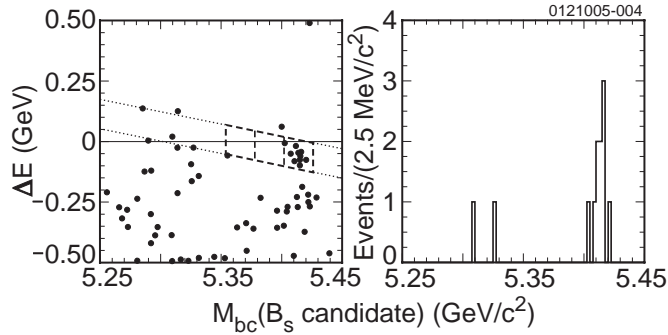,width=3.5in}
\end{center}
\caption{ The search plane~(left) and its projection on $M_{\rm bc}$~(right) for
          events in the signal band
          for $B_s$ modes with a $D_s^{(*)-}$ in the $\Y(5S)$ data.
	  The signal band (dotted lines) and signal boxes (dashed lines)
          are the same as in Figure~\ref{jpsi_y5s_signal}.
        }
\label{ds_y5s_data}
\end{figure}

Events satisfying the selection criteria in the $\Y(5S)$ data are 
shown in Figure~\ref{ds_y5s_data}~(left). There are 63 events 
in the search plane, 10 events are  in the signal region 
corresponding to $B_s^{*} \bar{B}_s^{*}$. 
Figure~\ref{ds_y5s_data}~(right) is a projection of the 
search plane on $M_{\rm bc}$ for events in the signal band. 
Table~\ref{dsbreakdown} shows the signal events by $\bar{B}_s$ and $D_s^+$ 
mode of reconstruction.

\begin{table}
\begin{center}
\begin{tabular}{ c c c c c c }
\hline
\hline
  & $D_s^+ \ra $   & $K^+ K^0_S$  & $K^+ \bar{K}^{0*}$  &
       $ \phi \pi^+$  &  $ \phi \rho^+$     \\
\hline
$\bar{B}_s \ra D_s^+     \pi^-/\rho^-$     &  &  0/0  & 1/1  & 1/3  & 1/1  \\
$\bar{B}_s \ra D_s^{*+}  \pi^-/\rho^-$     &  &  0/1  & 1/0  & 0/0  & 0/0  \\
\hline
\hline
\end{tabular}
\end{center}
\caption{ The $B_s$ candidates tabulated by $\bar{B}_s$ and $D_s^+$ mode.}
\label{dsbreakdown}
\end{table}

The probability, $P_{\rm II}$, for the background to fluctuate 
upwards and account for all events in the signal region for $B_s^* \bar{B}_s^*$ 
is estimated using the $\Y(5S)$ data in the sidebands of the signal region. 
To obtain a conservative estimate, we assume that the background is 
distributed uniformly over the 
lower half of the search plane. The number of background events in 
the signal region is less than 1.8 at 68\%~CL. The Poisson 
probability for 1.8 events to fluctuate to 10 or more events
in the signal region is $P_{\rm II} = 1.9 \times 10^{-5}$.

The probabilities $P_{\rm I}$ and $P_{\rm II}$ for the background 
to account for all events in the signal region for the two analyses 
are independent.
A combined probability $P$ is obtained as
$P = (P_{\rm I} P_{\rm II}) [1 - \ln( P_{\rm I} P_{\rm II} )]$~\cite{signif}.
We find $P = 7.7 \times 10^{-10}$, which corresponds
to a significance above 6.1$\sigma$~\cite{alternativeP}.

We calculate $\sigma(e^+ e^- \ra B_s^* \bar{B}_s^*)$ using 
$ \sigma(e^+ e^- \ra B_s^* \bar{B}_s^*) = 
      \frac{N_{\rm observed} - N_{\rm background}}
      {2 \mathcal{L} \epsilon^* } $,
where $\epsilon^*$ is the combined reconstruction efficiency
obtained using a GEANT-based Monte Carlo~(MC) simulation~\cite{GEANT}
for all modes including the $B_s$ and subsidiary branching 
fractions. The absolute reconstruction efficiencies 
range from a few per cent for $\bar{B}_s \rightarrow D_s^{*+} \rho^-$ 
to about 30\% for $B_s  \rightarrow J/\psi \phi$.
All $B_s$ branching fractions are unknown or 
poorly measured. We estimate the $B_s$ branching fractions by relating 
them to $B$ branching fractions that have contributions from the same 
quark-level diagrams, and assuming $SU(3)$ symmetry. 
For the $B_s \ra J/\psi \phi$, $B_s \ra J/\psi \eta$ and 
$B_s \ra J/\psi \eta^\prime$ modes, the following  branching 
fractions are used
$\mathcal{B}(B \ra J/\psi K^*) = (1.32 \pm 0.06) \times 10^{-3}$,
$\frac{1}{3} \mathcal{B}(B \ra J/\psi K) = (0.31 \pm 0.01) \times 10^{-3}$ 
and
$\frac{2}{3} \mathcal{B}(B \ra J/\psi K) = (0.63 \pm 0.02) \times 10^{-3}$,
respectively. For the $\bar{B}_s \ra D_s^{(*)+} \pi^-/\rho^-$  
modes, only the corresponding $\bar{B}^0$ branching fractions,
$i.e.,$ $\mathcal{B}( \bar{B}^0 \ra D^{(*)+} \pi^-/\rho^-) $, 
are used, as they proceed predominantly through an external 
spectator diagram.
For $\mathcal{B}(D_s^+ \ra \phi \pi^+)$ 
a weighted average  of the PDG average~\cite{pdg} and a recent 
measurement~\cite{dsphipibabar} is used.  The $D_s^+$  
branching fractions for the other  three modes are updated 
accordingly, as they are all measured with respect to 
$D_s^+ \ra \phi \pi^+$. Other subsidiary branching fractions 
are well known~\cite{pdg}.

Most of the data was taken at $E_{\rm CM} = 10.859$~GeV,
however, a small subset, which contains one signal event,  
was taken at an energy about 56 MeV higher. In order to 
quote the cross section at the $\Y(5S)$ peak we exclude 
this event from the signal yield. Using the remaining 
13 $B_s$ candidates, we find:
$\sigma(e^+ e^- \ra B_s^* \bar{B}_s^*)  = 
[0.11^{+0.04}_{-0.03}  {\rm (stat.)} \; \pm 0.02 {\rm (syst.)}]\;{\rm nb}.$
The systematic uncertainty has large contributions 
from the uncertainties in $B$ and $D_s^+$ branching 
fractions, and the assumption of $SU(3)$ symmetry. 
Other uncertainties in  track, $\pi^0$, and $K_S^0$ finding 
efficiencies, particle identification efficiencies, 
the background estimates and the integrated luminosity of 
the $\Y(5S)$ data sample are small.

The number of events consistent with $B_s^{*} \bar{B}_s^{*}$ 
production at $E_{\rm CM} = 10.859$~GeV is 13, while the 
number of events consistent with either of the other two 
$B_s$ production mechanisms is 0. 
Accounting for the background in the signal region, we find:
$\sigma(e^+ e^- \ra B_s \bar{B}_s) / 
 \sigma(e^+ e^- \ra B_s^{*} \bar{B}_s^{*}) < 0.16 $ 
and 
$[ \sigma(e^+ e^- \ra B_s \bar{B}_s^{*}) +  
   \sigma(e^+ e^- \ra B_s^{*} \bar{B}_s ) ] /
   \sigma( e^+ e^- \ra B_s^{*} \bar{B}_s^{*} ) < 0.16 $ 
at 90\%~CL.

Using all 14 signal $B_s$ candidates, the mass of the $B_s^*$ meson
is measured to be $M_{B_s^*} = 
[5.414 \pm 0.001{\rm (stat.)} \pm 0.003{\rm (syst.)} ]~{\rm GeV}/c^2$.
The dominant systematic uncertainty arises from imperfect 
knowledge of the absolute beam energy scale~(2.9~MeV/$c^2$), which was calibrated 
using data collected at the narrow $\Y(1S)$, $\Y(2S)$ and $\Y(3S)$ 
resonances, as well as data collected at the $\Y(4S)$ resonance. 
Using the $B_s$ mass measurement in Ref.~\cite{bsmass}, we also find:
$ M_{B_s^*} - M_{B_s}  = [48 \pm 1 {(\rm stat.)} \pm 3 {(\rm syst.)}]~{\rm MeV}/c^2$,
which is consistent with an earlier measurement
of $M_{B_s^*} -  M_{B_s}= (47.0 \pm 2.6)$~MeV/$c^2$~\cite{pdg}.

In summary, using the CLEO detector at CESR,  
we have observed the $B_s$ meson in $e^+ e^-$ annihilation at the $\Y(5S)$ 
resonance. We have established that $B_s$ meson production proceeds 
predominantly through the creation of $B_s^{*} \bar{B}_s^{*}$ pairs. 
We find $\sigma(e^+ e^- \ra B_s^* \bar{B}_s^*)  
= [0.11^{+0.04}_{-0.03}{\rm (stat.)} \pm 0.02 {\rm (syst.)}]  \; {\rm nb}$, 
and set the following limits: $ \sigma(e^+ e^- \ra B_s \bar{B}_s) /
\sigma(e^+ e^- \ra B_s^{*} \bar{B}_s^{*}) < 0.16 $ and 
$[ \sigma(e^+ e^- \ra B_s \bar{B}_s^{*}) + \sigma(e^+ e^- \ra B_s^{*} \bar{B}_s) ] 
/  \sigma(e^+ e^- \ra B_s^{*} \bar{B}_s^{*} ) < 0.16$ at 90\%~CL.
The observation that $B_s$ pairs are produced predominantly 
in the $B_s^* \bar{B}_s^*$ configuration is in agreement with 
the prediction of the Unitarized Quark Model~\cite{theory}
and predictions in Ref.~\cite{rosner}. 
The mass of the $B_s^*$ meson is measured to be 
$M_{B_s^*} = [ 5.414 \pm 0.001{\rm (stat.)} \pm 
                          0.003{\rm (syst.)} ]$~GeV/$c^2$.

We gratefully acknowledge the effort of the CESR staff in
providing us with excellent luminosity and running conditions.
This work was supported by the National Science Foundation and the
U.S. Department of Energy.


\begin{thebibliography}{99}
\bibitem{alternativeP}
 An alternative estimate of $P$ can be obtained from the probability 
 for the total number of background events in the two analyses~(1.9 events) 
 to fluctuate to the number of signal candidates~(14 events): 
 $P = 1.6\times 10^{-8}$~(5.7$\sigma$).
\end{thebibliography}
\end{document}